\documentclass[twocolumn,english,aps,prl,superscriptaddress,groupedaddress]{revtex4}
\usepackage[utf8]{inputenc}
\usepackage{amsmath}
\usepackage{amssymb}
\usepackage{graphicx}
\usepackage{esint}

\makeatletter
\@ifundefined{textcolor}{}
{%
 \definecolor{BLACK}{gray}{0}
 \definecolor{WHITE}{gray}{1}
 \definecolor{RED}{rgb}{1,0,0}
 \definecolor{GREEN}{rgb}{0,1,0}
 \definecolor{BLUE}{rgb}{0,0,1}
 \definecolor{CYAN}{cmyk}{1,0,0,0}
 \definecolor{MAGENTA}{cmyk}{0,1,0,0}
 \definecolor{YELLOW}{cmyk}{0,0,1,0}
}

\usepackage{epsfig}
\usepackage{epstopdf}
\usepackage{bm}
\usepackage{epsfig}\usepackage{mathrsfs}
\usepackage{dcolumn}
\usepackage{color}
\usepackage{miniplot}
\usepackage{minitoc}
\usepackage{blindtext}
\@ifundefined{showcaptionsetup}{}{%
\PassOptionsToPackage{caption=false}{subfig}}
\def\be{\begin{equation}}
\def\ee{\end{equation}}
\def\bea{\begin{eqnarray}}
\def\eea{\end{eqnarray}}
\def\la{\langle}
\def\ra{\rangle}

\allowdisplaybreaks[2]
\DeclareUnicodeCharacter{00A0}{~}

\usepackage{babel}

\usepackage{babel}

\makeatother

\usepackage{babel}
\begin{document}
\global\long\def\ket#1{\left| #1\right\rangle }

\global\long\def\bra#1{\left\langle #1\right|}

\global\long\def\braket#1#2{\left\langle #1| #2\right\rangle }

\title{Observation of momentum-space chiral edge currents in room-temperature atoms }

\author{Han Cai}

\affiliation{Interdisciplinary Center of Quantum Information and Department of
Physics, Zhejiang University, Hangzhou 310027, People’s Republic of
China}

\author{Jinhong Liu}

\affiliation{Interdisciplinary Center of Quantum Information and Department of
Physics, Zhejiang University, Hangzhou 310027, People’s Republic of
China}

\affiliation{State Key Laboratory of Quantum Optics and Quantum Optics Devices,
Institute of Opto-Electronics, Shanxi University, Taiyuan 030006,
People’s Republic of China}

\author{Jinze Wu}

\affiliation{Interdisciplinary Center of Quantum Information and Department of
Physics, Zhejiang University, Hangzhou 310027, People’s Republic of
China}

\affiliation{State Key Laboratory of Quantum Optics and Quantum Optics Devices,
Institute of Opto-Electronics, Shanxi University, Taiyuan 030006,
People’s Republic of China}

\author{Yanyan He}

\affiliation{Interdisciplinary Center of Quantum Information and Department of
Physics, Zhejiang University, Hangzhou 310027, People’s Republic of
China}

\author{Shi-Yao Zhu}

\affiliation{Interdisciplinary Center of Quantum Information and Department of
Physics, Zhejiang University, Hangzhou 310027, People’s Republic of
China}

\affiliation{Synergetic Innovation Center of Quantum Information and Quantum Physics,
University of Science and Technology of China, Hefei, Anhui 230026,
People’s Republic of China}

\author{Jun-Xiang Zhang}
\email{junxiangzhang@zju.edu.cn}

\selectlanguage{english}%

\affiliation{Interdisciplinary Center of Quantum Information and Department of
Physics, Zhejiang University, Hangzhou 310027, People’s Republic of
China}

\author{Da-Wei Wang}
\email{dwwang@zju.edu.cn}

\selectlanguage{english}%

\affiliation{Interdisciplinary Center of Quantum Information and Department of
Physics, Zhejiang University, Hangzhou 310027, People’s Republic of
China}

\affiliation{CAS Center of Excellence in Topological Quantum Computation, University
of Chinese Academy of Sciences, Beijing 100190, People’s Republic
of China}

\date{\today }
\begin{abstract}
Chiral edge currents play an important role in characterizing topological matter. In atoms, they have been observed at such a low temperature that the atomic motion can be measured. Here we report the first experimental observation of chiral edge currents in atoms at room temperature. Staggered magnetic fluxes are induced by the spatial phase difference between two standing-wave light fields, which couple atoms to form a momentum-space zigzag superradiance lattice. The chiral edge currents have been measured by comparing the directional superradiant emissions of two timed Dicke states in the lattice. This work paves the way for quantum simulation of topological matter with hot atoms and facilitates the application of topological physics in real devices. 
\end{abstract}


\maketitle
The quantum Hall effect \cite{Klitzing1980} reveals a topological
class of matter that are charaterized by the Chern numbers of energy
bands \cite{Thouless1982}. The chiral edge currents located at the
boundaries of two bulk materials with different Chern numbers are usually
measured to investigate the band topology. The chirality of the edge currents
is featured by the locking between the direction of the currents and
the (pseudo-)spin states of the edge excitations \cite{Kane2005,Bernevig2006}.
The chirality is robust against local perturbations and only
changes when the energy bands go through a topological transition.
Since the edges have a lower dimension than the bulk, the edge currents
provide a convenient platform to investigate topological physics in
a higher dimension, such as the quantum Hall effect in four dimensions
\cite{Lohse2018,Zilberberg2018}. The chirality of the edge currents
persists even when a two-dimensional lattice is reduced to quasi one-dimensional
ribbons \cite{HuegelParedes2014}, which has been experimentally demonstrated
for ultracold fermions \cite{Mancini2015,Livi2016} and bosons \cite{Atala2014,Stuhl2015}.
In those experiments, the chiral edge currents were measured with
the atomic motions and thus were only observed at a low temperature,
where the thermal motions are negligible.

Here we report the experimental observation of the momentum-space
chiral edge currents in a superradiance lattice \cite{Wang2015a,Wang2015b,Chen2018}
of cesium atoms at room temperature. The zigzag lattice that we have
synthesized is similar to the ladder structures in the experiments
with cold atoms \cite{Mancini2015,Stuhl2015,HuegelParedes2014} and
is currently under intensive investigation \cite{Anisimovas2016,Xu2018,An2018}.
Different from the momentum-space lattices characterized by the recoil
momentum of cold atoms \cite{An2017,An2018}, the superradiance lattice
is a momentum-space lattice composed by the timed Dicke states \cite{Scully2006},
which are collective atomic excitations with phase correlations. The
phase correlations can be understood as the momenta of the collective
excitations, which have directional superradiant light emissions when
they satisfy the phase-matching condition with a light mode \cite{Scully2006}.
A remarkable advantage of our approach is that the edge currents are
observed at room temperature. Instead of measuring the atomic motions
\cite{Mancini2015,Livi2016,Atala2014,Stuhl2015}, the chiral edge
currents are measured by comparing the directional light emissions
from two timed Dicke states. Strikingly, the chiral edge currents are robust to 
room-temperature thermal motions, which
only induce an average effect of the chiral edge currents in the real-space
Brillouin zone. Our study has
substantially lowered the threshold of the experimental observation
of chiral edge currents in atoms.

\begin{figure}
\includegraphics[width=1\columnwidth]{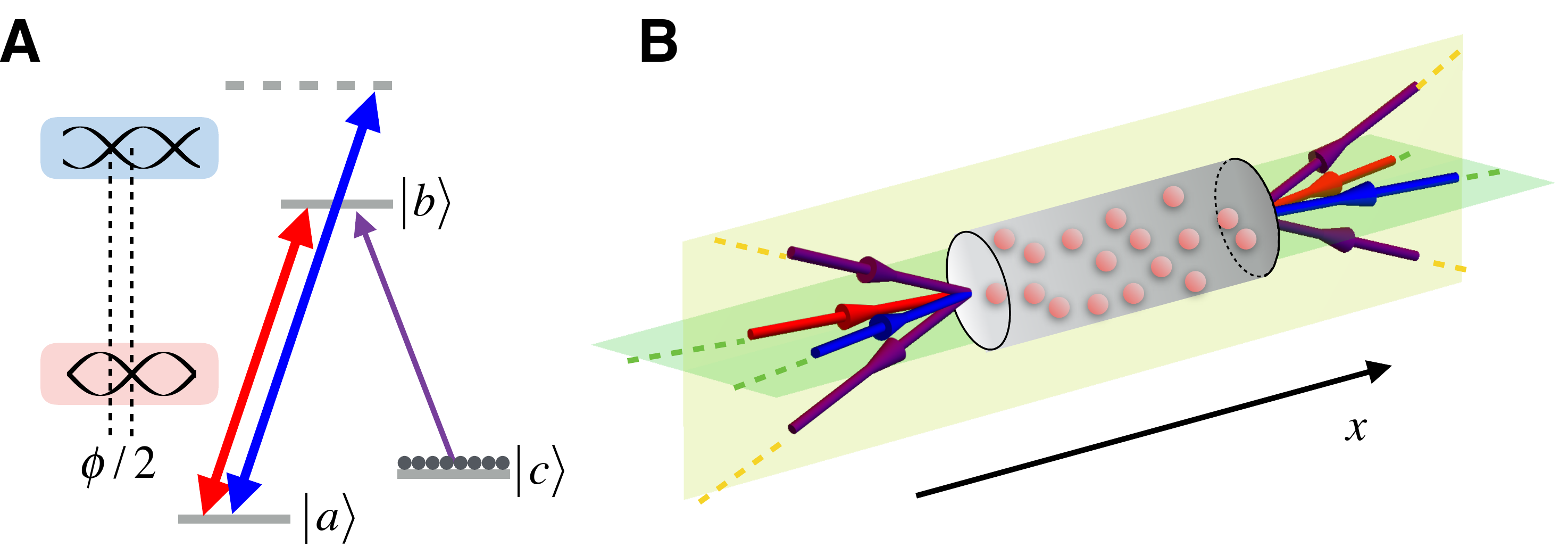}\caption{\textbf{Schematic configuration of the experiment}. 
(\textbf{A}) Atomic levels and the near-resonant and far-detuned standing waves coupling
fields with a relative spatial phase difference $\phi/2$. (\textbf{B})
The configuration of the lasers. Both the far-detuned (blue) and near-resonant
(red) standing wave coupling fields have an angle to the $\hat{x}$-axis 
in order to satisfy the phase-matching condition.
The plane formed by the two standing wave coupling light beams is
perpendicular to the one formed by the incident and reflected probe
beams.}
\label{fig1} 
\end{figure}

To highlight the physics, we introduce our basic model with $\Lambda$-type
three-level atoms as shown in Fig. \ref{fig1} (\textbf{A}).
An excited state $|b\ra$ and a metastable state $|a\ra$ are coupled
by two standing waves with different frequencies, i.e., a near-resonant
and a far-detuned standing waves with field amplitude envelopes
$\cos(k_{c}x)$ and $\cos(k_{c}x+\phi/2)$, where $\phi/2$ is their
spatial phase difference. The detuning between the two standing wave
coupling fields is small enough such that the difference between the
amplitudes of their wavevectors $k_{c}$ can be neglected. A probe
field couples the ground state $|c\rangle$ to $|b\rangle$. The Hamiltonian
is (we set $\hbar=1$), 
\begin{equation}
\begin{aligned}H= & \sum_{m}[\frac{\Delta_{c}}{2}+2\kappa\cos(2k_{c}x+\phi)](|a_{m}\ra\la a_{m}|-|b_{m}\ra\la b_{m}|)\\
 & +\sum_{m}2\Omega\cos(k_{c}{x}_{m})(|b_{m}\ra\la a_{m}|+h.c.)\\
 & +\sum_{m}[\Omega_{p}e^{i{k}_{p}x_{m}}e^{-i[\Delta_{p}-(\nu_{c}-\omega_{ba})/2]t}|b_{m}\ra\la c_{m}|+h.c.],
\end{aligned}
\label{HEIT}
\end{equation}
where $\Delta_{c}=\nu_{c}-\omega_{ba}+4\kappa$ is the detuning of
the near-resonant coupling field frequency $\nu_{c}$ from the atomic
transition between $|b\rangle$ and $|a\rangle$. $\omega_{ba}$ is
the bare transition frequency and $4\kappa$ is a spatially homogeneous
Stark shift induced by the far-detuned coupling fields, which also
induce a spatially periodic Stark shift $2\kappa\cos(2k_{c}x+\phi)$.
Here $\kappa=\Omega_{f}^{2}/\Delta_{f}$ with $\Omega_{f}$ and $\Delta_{f}=\nu_{f}-\omega_{ba}$
being the Rabi frequency and detuning of each plane wave component
of the far-detuned coupling field. $\Delta_{p}=\nu_{p}-\omega_{bc}$
is the detuning of the probe field frequency $\nu_{p}$ from the atomic
transition frequency $\omega_{bc}$ between $|b\rangle$ and $|c\rangle$.
$\Omega$ is the Rabi frequency of each plane wave component of the
near-resonant coupling fields. ${k}_{p(c)}$ is the $\hat{x}$-component
of the probe (coupling) field wavevector and ${x}_{m}$ is the $\hat{x}$-axis
coordinate of the $m^{\textrm{th}}$ atom. The derivation of the Hamiltonian in Eq.~(\ref{HEIT})
can be found in Sec.~I of the Supplementary Material.

\begin{figure}
\includegraphics[width=1\columnwidth]{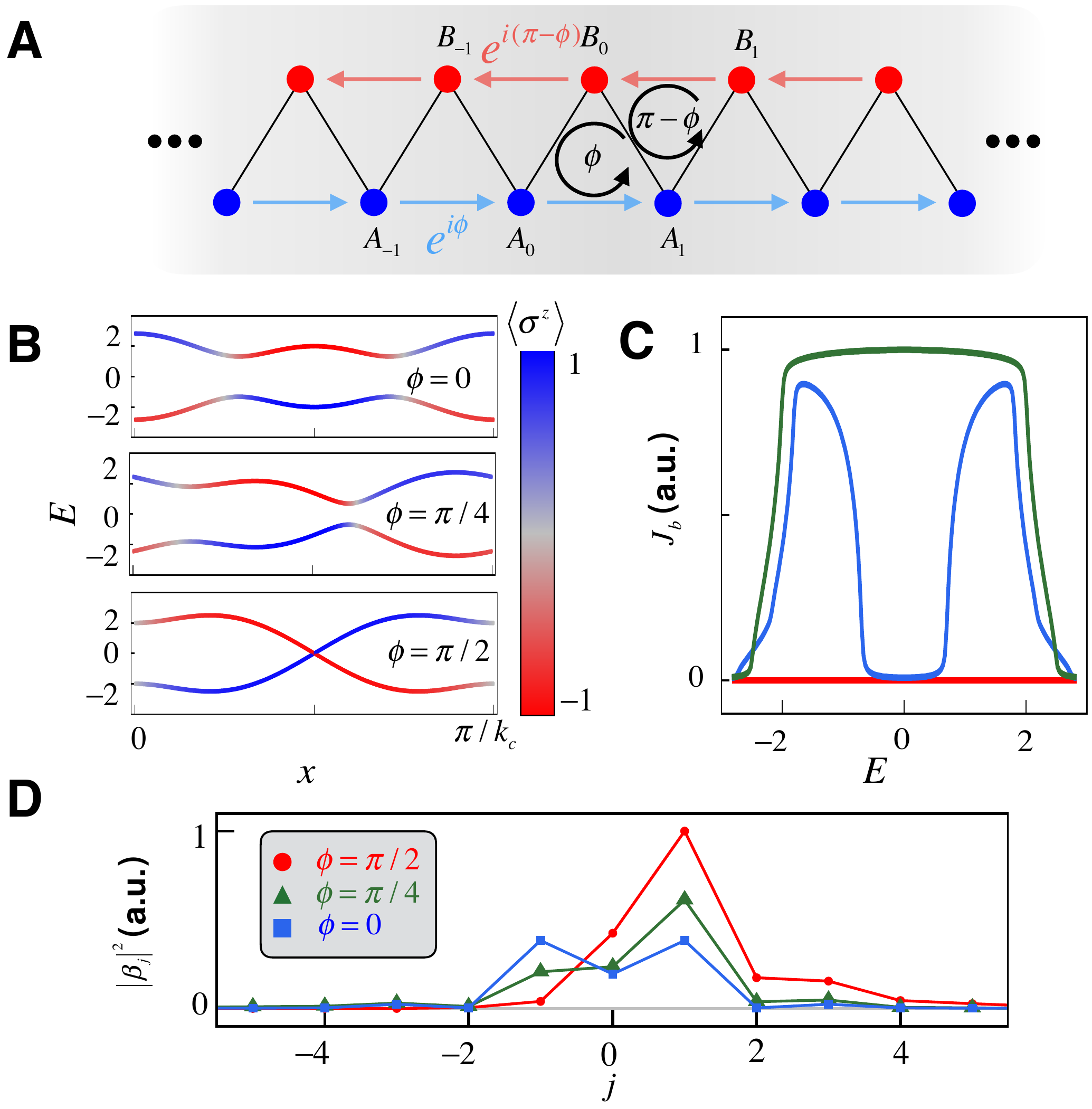}\caption{\textbf{The 
superradiance lattice and its band structure with different magnetic
fluxes.} (\textbf{A}) The tight-binding lattice of the Hamiltonian
$H_{s}$ in Eq. (\ref{H}). The red and blue arrows show the phase
factors $\phi$ attached with the corresponding transitions. The total
phases enclosed by the loop transitions in the down and up triangles
are $\phi$ and $\pi-\phi$, respectively. The relative phase $\phi=\pi/2$.
(\textbf{B}) The dispersion relation according to Eq.~(\ref{Hk})
with $\phi=0,\pi/4$ and $\pi/2$. The colour shows the $\langle \sigma_z\rangle$ of
eigenstates and indicates which edge the eigenstates 
are mainly located on. (\textbf{C}) $J_b$ in Eq. (\ref{Jb}) 
for $\phi=0$ (blue square), $\pi/4$ (green triangle) and $\pi/2$ (red circle).
(\textbf{D}) The steady state
distribution of the population in the $|b\rangle$-sublattice with a probe field pumping
the atoms from the ground state to the state $|B_{0}\rangle$. The
decoherence rates of states $|b\rangle$ and $|a\rangle$ are $\gamma_{bc}=1$
and $\gamma_{ac}=0.1$. The probe detuning $\Delta_{p}=0$. $\Omega=1$, 
$\kappa=1$ and $\Delta_c=0$.}
\label{fig2} 
\end{figure}

By introducing collective atomic excitation operators $\hat{b}_{j}^{\dagger}=1/\sqrt{N}\sum_{m}e^{i({k}_{p}+2j{k}_{c}){x}_{m}}|b_{m}\ra\la c_{m}|$
and $\hat{a}_{j}^{\dagger}=1/\sqrt{N}\sum_{m}e^{i[{k}_{p}+(2j-1){k}_{c}]{x}_{m}}|a_{m}\ra\la c_{m}|$,
we transform the Hamiltonian to momentum space, $H=H_{s}+H_{p}$ where
\begin{eqnarray}
\begin{aligned}H_{s}= & \sum\limits _{j}\Delta_{c}(\hat{a}_{j}^{\dagger}\hat{a}_{j}-\hat{b}_{j}^{\dagger}\hat{b}_{j})/2+\sum\limits _{j}[\Omega(\hat{a}_{j}^{\dagger}\hat{b}_{j}+\hat{a}_{j}^{\dagger}\hat{b}_{j-1})\\
 & +e^{i\phi}\kappa(\hat{a}_{j}^{\dagger}\hat{a}_{j-1}-\hat{b}_{j}^{\dagger}\hat{b}_{j-1})+h.c.],
\end{aligned}
\label{H}
\end{eqnarray}
and $H_{p}=\sqrt{N}\Omega_{p}[\hat{b}_{0}^{\dagger}e^{-i[\Delta_{p}-(\nu_{c}-\omega_{ba})/2]t}+h.c.]$.
With the condition that $\Omega\gg\Omega_{p}$, most of the atoms
are in the ground state $|c\rangle$, and $\hat{a}_{j}^{\dagger}$
and $\hat{b}_{j}^{\dagger}$ are bosonic \cite{Fleischhauer2000,Wang2015a}.
For single excitations, Eq. (\ref{H}) is a Hamiltonian of a tight-binding
superradiance lattice composed by timed Dicke states, $|B_{j}\ra\equiv\hat{b}_{j}^{\dagger}|c_{1},c_{2},...,c_{N}\ra$
and $|A_{j}\ra\equiv\hat{a}_{j}^{\dagger}|c_{1},c_{2},...,c_{N}\ra$.
For more excitations, as long as the excitation number is much less
than the atomic number, the physics remains the same due to the bosonic
nature of the excitations.

\begin{figure*}
\includegraphics[width=2\columnwidth]{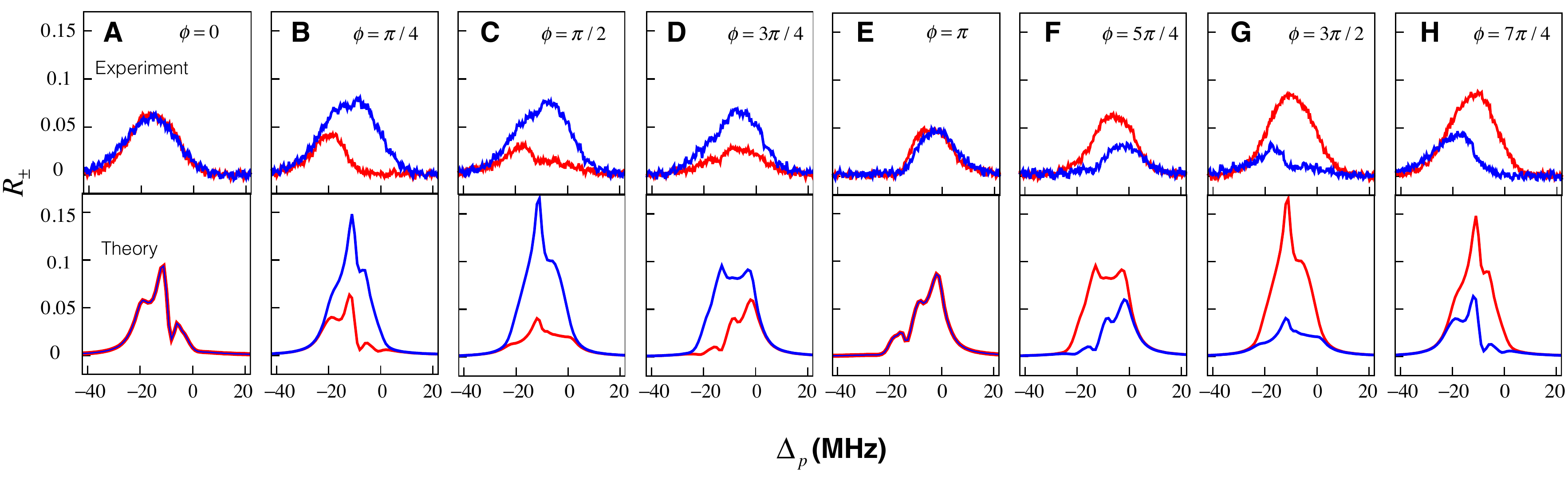} \caption{\textbf{The chiral edge 
currents demonstrated by the reflection spectra.} 
$R_{+}$ are red lines and $R_{-}$ are blue lines with
phases (\textbf{A}) $\phi=0$, (\textbf{B}) $\phi=\pi/4$,
(\textbf{C}) $\phi=\pi/2$, (\textbf{D}) $\phi=3\pi/4$,
(\textbf{E}) $\phi=\pi$, (\textbf{F}) $\phi=5\pi/4$,
(\textbf{G}) $\phi=3\pi/2$ and (\textbf{H}) $\phi=7\pi/4$.
The power of each plane wave component of the near-resonant coupling
field is 6.5 mW with an effective Rabi frequency $\Omega=5.5$ MHz.
The power of each plane wave component of the far-detuned coupling
field is 120 mW with an effective Rabi frequency $\Omega_{f}=33.3$
MHz and a detuning $\Delta_{f}=200$ MHz. $\kappa=5.5$ MHz.
$\Delta_c=1$ MHz. The power of each probe beam is 20 $\mu$W. }
\label{fig3} 
\end{figure*}

The phase factor $e^{i\phi}$ in Eq. (\ref{H}) is induced by the
spatial phase difference between the two standing wave coupling fields.
Each up and down triangle encloses an effective magnetic flux $\phi$
and $\pi-\phi$, respectively, as shown in Fig. \ref{fig2} (\textbf{A}).
A direct consequence of this effective magnetic field is the chiral
edge currents, which can be demonstrated by the dispersion relation
of $H_{s}$. We diagonalize $H_{s}$ in real space, 
\begin{equation}
H_{s}=h\mathbf{n}\cdot\mathbf{\sigma},\label{Hk}
\end{equation}
where $\mathbf{n}=(h_{x}\hat{x}+h_{z}\hat{z})/h$ with $h_{x}=2\Omega\cos({k}_{c}{x})$,
$h_{z}=\Delta_{c}/2+2\kappa\cos(2{k}_{c}{x}+\phi)$, and $h=\sqrt{h_{z}^{2}+h_{x}^{2}}$.
$\mathbf{\sigma}=\sum_{j=x,y,z}\sigma^{j}\hat{j}$ is the vector of
the Pauli matrices of the pseudo spin-up state $|a\rangle$ and spin-down
state $|b\rangle$. The dispersion relations in the two bands are
$E_\pm=\pm h$ with the eigenstates $|\psi_{+}\ra=\cos(\theta/2)|a\ra+\sin(\theta/2)|b\ra$
and $|\psi_{-}\ra=-\sin(\theta/2)|a\ra+\cos(\theta/2)|b\ra$, where
$\theta$ is the polar angle of $\mathbf{n}$. In Fig. \ref{fig2}
(\textbf{B}), we plot the dispersion with the ``spin texture''
$\left\langle \sigma^{z}\right\rangle $. For $\phi=\pi/2$, 
most eigenstates concentrate on one of the two edges. 

The evolution of the momentum is determined by the dispersion relation in Fig. \ref{fig2} (\textbf{B}),
$\partial p_\pm/\partial t=-\partial E_\pm/\partial x$ \cite {Wang2015a}.
For example, when $\phi=\pi/2$ and near the energy $E=0$, the momentum of an excitation created on the $|b\rangle$-sublattice
increases with time (note the negative derivative of the red line), while on $|a\rangle$-sublattices the momentum decreases.
This is the essence of chiral edge currents, i.e., excitations on different edges or with different
spin states move in opposite directions. To quantitatively clarify this feature, we define
the chiral edge currents on the $|b\rangle$-sublattice as \cite{HuegelParedes2014},
\begin{equation}
J_{b}(E)=-\sum_{i=\pm} \int \text{d}x \delta{(E_i-E)} |\langle\psi_i|b\rangle|^2 \frac{\partial E_i}{\partial x},
\label{Jb}
\end{equation} 
where $\delta(E_i-E)$ is the Dirac delta function. $J_b(E)$ characterizes the dynamics of the total momentum of excitations with energy $E$ on the
$|b\rangle$-sublattice. We can define a similar quantity for the $|a\rangle$-sublattice. 
However, we only focus on $J_b(E)$ since the probe field couples atoms to the $|b\rangle$ state.
In Fig. \ref{fig2} (\textbf{C}), we plot $J_b(E)$ for different phases $\phi$.
For $\phi=0$, $J_b(E)=0$. For $\phi=\pi/4$, $J_b(E)$
is positive in the two bands. $J_b(E)$ is almost a big positive constant when $\phi=\pi/2$.
When $\phi=\pi$, $J_b(E)=0$ and when $\phi=3\pi/2$, $J_b(E)$ is negative
to the one with $\phi=\pi/2$ (see Sec. II of the Supplementary Material).

To generate $J_b(E)$, we apply a weak probe field with detuning $\Delta_p=E+(\nu_c-\omega_{ba})/2$ to pump the atoms from the ground state
to the state $|B_{0}\rangle$. The excitation induces $J_b(E)$ and is finally balanced by the decoherence
of the atomic states. In the steady state, the population distribution
is plotted in Fig. \ref{fig2} (\textbf{D}). We define $\beta_{j}$
as the probability amplitude of the state $|B_{j}\rangle$. For $\phi=0$,
the distribution of $|\beta_{j}|^{2}$ is symmetric on both sides
of $|B_{0}\rangle$, in contrasted to the asymmetric distribution when $\phi=\pi/4$
and $\phi=\pi/2$, where the population is biased to $j>0$. The result is consistent with Fig. \ref{fig2} (\textbf{C}).

In the experiment, we detect the superradiant emissions of two specific
timed Dicke states to show the edge currents. The timed Dicke states
with a phase correlation that matches the wavevectors of the light
in the medium, i.e., $|{k}_{p}+2j{k}_{c}|\approx|{k}_{p}|$, have
directional superradiant emissions \cite{Scully2006}. In the current
scheme where $|a\rangle$ and $|c\rangle$ are nearly degenerate,
the timed Dicke state $|B_{-1}\rangle$ (or $|B_{+1}\rangle$) is
the only superradiant state besides $|B_{0}\rangle$ for a probe light
incident along $+\hat{x}$ (or $-\hat{x}$). The radiation from these
two superradiant states can be considered as the reflectivities $R_{\pm}$
of the probe fields incident to the atoms along $\pm\hat{x}$. The
relationship between $R_{\pm}$ and $\beta_{\pm1}$ is (see Methods
and a rigorous calculation in Sec.~III of the Supplementary Material), 
\begin{equation}
\frac{R_{+}}{R_{-}}=\frac{|\beta_{-1}|^{2}}{|\beta_{+1}|^{2}}.\label{eq:ratio}
\end{equation}
This relation is independent of the density of the atoms, the length
of the vapor cell and the phase mismatch.

We have used the D1 line of cesium atoms in the experiment: $|a\ra=|6^{2}S_{1/2},F=3\ra$,
$|b\ra=|6^{2}P_{1/2},F=4\ra$ and $|c\ra=|6^{2}S_{1/2},F=4\ra$ (see Methods for details of the experiment). Typical
experimental results are shown in Fig. \ref{fig3}. The reflection
spectra depend on the phase $\phi$. Only when $\phi=0$ and $\pi$,
we observe $R_{+}=R_{-}$, as shown in Figs. \ref{fig3} (\textbf{A})
and (\textbf{E}). For $0<\phi<\pi$, we observe $R_{+}<R_{-}$, as shown
in Figs. \ref{fig3} (\textbf{B})-(\textbf{D}), which indicates a larger
population in $|B_{+1}\rangle$ than in $|B_{-1}\rangle$, resulted
from an edge current propagating along $+\hat{x}$. This is consistent
with the results in Figs. \ref{fig2} (\textbf{C}) and (\textbf{D}). In contrast,
we observe $R_{+}>R_{-}$ for $\pi<\phi<2\pi$, as shown in Figs. \ref{fig3}
(\text{\textbf F})-(\textbf{H}), which indicates an edge current along $-\hat{x}$.
The differences between the simulation and experiment are attributed
to slight asymmetry in the optical alignment, an average of $\phi$
along the vapour cell ($\Delta\phi\approx 0.05\pi$) and the Gaussian rather than plane wave profiles
of the coupling fields.

\begin{figure}
\includegraphics[width=1\columnwidth]{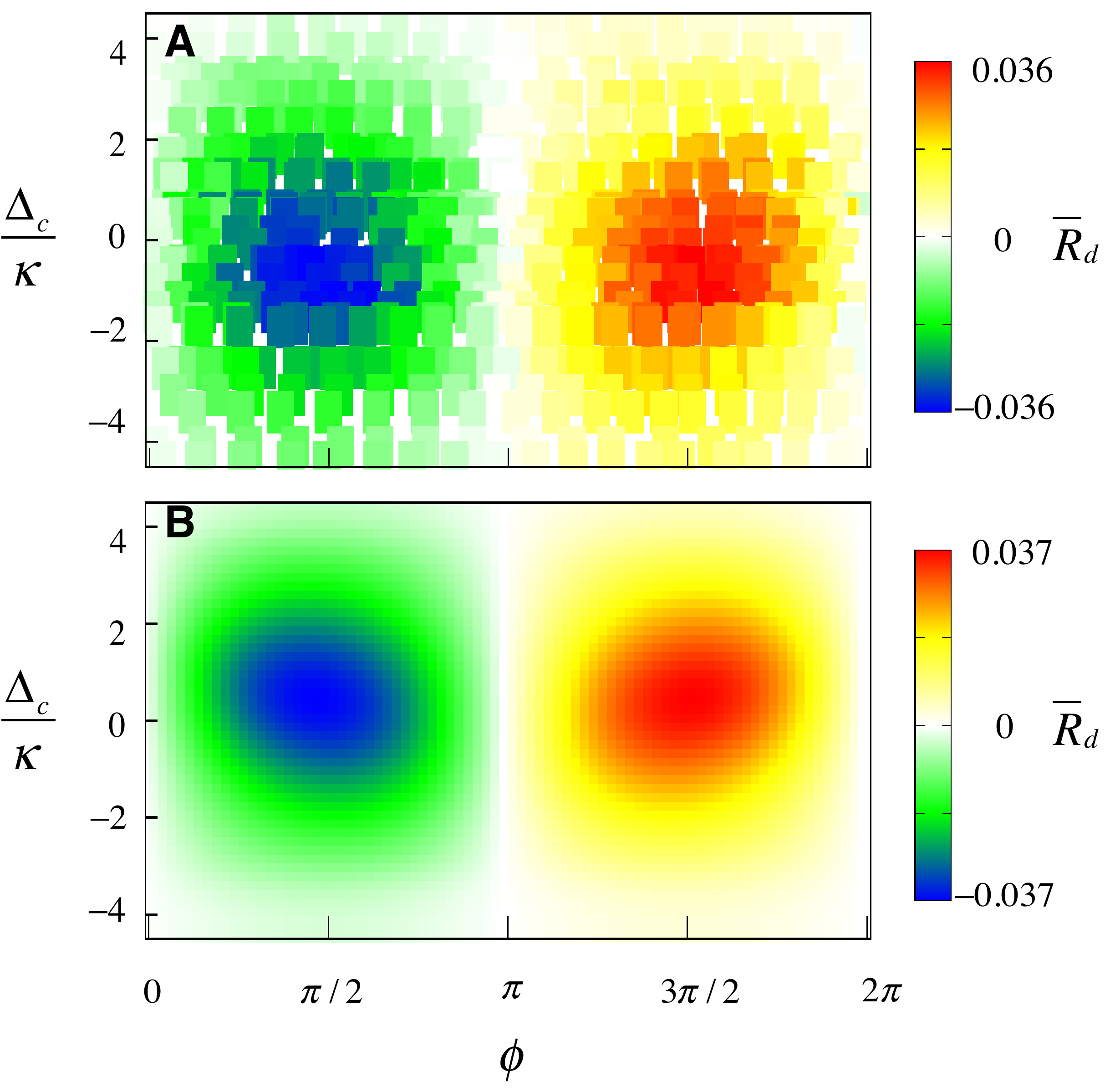}\caption{\textbf{The averaged difference between the two reflectivities in Eq. (\ref{Rd}).}
(\textbf{A}) Experimental data. (\textbf{B}) Numerical
simulation. The parameters are the same as in
Fig.~\ref{fig3}.}
\label{fig4} 
\end{figure}

In the following, we analyse the robustness of the edge currents against
thermal motions. From
Fig. \ref{fig3}, we see that the reflection spectra are not Doppler
broadened. The non-zero region of the spectra coincides with the energy
bands of the superradiance lattice. Their scales are both around 30 MHz. 
This feature of standing-wave coupled electromagnetically 
induced transparency was already found by Feldman
and Feld in 1972 \cite{Feldman1972}. To understand this, we notice
that the atoms that have a Doppler shift larger than the bandwidth
is out of resonance with the probe field no matter their positions such that they
cannot be excited. For atoms with Doppler shift smaller than the lattice
bandwidth, they move in the real-space Brillouin zone and their contribution
to the edge currents needs to be averaged with their positions. The
chiral edge currents induce a difference between the populations on
$|B_{-1}\rangle$ and $|B_{+1}\rangle$, and subsequently a difference
between the two reflectivities $R_{d}\equiv R_{+}-R_{-}\propto\left|\left\langle \beta_{-1}\right\rangle \right|^{2}-\left|\left\langle \beta_{+1}\right\rangle \right|^{2}$
where $\left\langle \right\rangle $ is the average over the Doppler
shifts due to the thermal motions of the atoms. To quantify the effect of the chiral edge currents
on the whole reflection spectra, we make average of $R_{d}$, 
\begin{equation}
\bar{R}_{d}=\frac{1}{F}\int R_{d}\text{d}\Delta_{p},\label{Rd}
\end{equation}
where $F=40$ MHz is the frequency range of an integration from $-30$
MHz to $10$ MHz in the reflection spectra. In Fig. \ref{fig4}, we
show the experimental data and numerical simulation of $\bar{R}_{d}$
as functions of $\Delta_c$ and $\phi$. For an on-site potential difference $\Delta_c\approx 0$, $\bar{R}_d$
is approximately a sinusoidal function of the phase $\phi$. This means our method
can be used to measure a phase difference between two standing waves.
For $\Delta_c$ larger than the bandwidth, 
the inter-edge transitions are inhibited and the effect of the synthetic magnetic field
diminishes due to the inefficient loop transitions. As a result, $\bar{R}_d$ decreases rapidly to zero when $\Delta_c$ increases.
The results in Fig. \ref{fig4} can be understood as the phase diagram of
an extended Haldane model \cite{HuegelParedes2014} and its relation to the dynamic classification
of topological phases \cite{Zhang2018} will be discussed elsewhere.

The results reported here are substantially different from the temperature-independent
edge currents in the photonic lattices \cite{Rechtsman2013,Hafezi2013,Zhang2015,Xue2017,Bandres2018},
where the propagation of the photons governed by the Maxwell equations
is made analogy to the Schrödinger equations \cite{Haldane2008}.
The edge states there are photonic states rather than atomic states.
The temperature has no influence on the photons. In our current study,
the topological bands are for the atoms and they intrinsically obey
the Schrödinger equation. The thermal motions of the atoms make a
convenient average of the edge currents. In addition, our
lattice is in momentum space, in contrast to the real-space topological
photonic lattices. Although the edge currents are detected by light
in our experiment, they are currents of collective excitations of
atoms in momentum space, not light in real space.

On the other hand, our results are closely related to the spin-orbit
coupled system \cite{Wu2016,Huang2016} and the momentum-space lattice \cite{An2017,An2018}
in cold atoms, with the difference that the momentum is represented
by the phase correlation of the collective excitation, instead of
the recoil momentum, which is negligible in our study. 
An extension of our model to higher dimensions \cite{Wang2015b}
can be used to simulate the Haldane model \cite{Haldane1988,Jotzu2014} and the two-dimensional
spin-orbit coupling \cite{Wu2016,Huang2016}. By using Rydberg states \cite{Peyronel2012,Ripka2016}
we can introduce interactions between the excitations and study the
many-body effect in flux lattices \cite{Greschner2017}. An interesting
connection of our results can also be made to the unidirectional reflectionless (invisible)
photonic structures \cite{Lin2011b,Regensburger2012,Wu2014,Huang2017}, such as the parity-time symmetric materials \cite{Feng2017}. We have observed
that under certain conditions one of the two reflectivities is nearly
zero while the other is big. Another observation is that the transmission
of the probe fields in the two opposite directions are the same, although
the transmissions are phase-dependent. This is because our system
does not break the time-reversal symmetry. It is interesting to note
that an effective magnetic field in momentum space does not result
in optical nonreciprocity, while an effective electric field in momentum
space can break the time-reversal symmetry and induce optical nonreciprocity
\cite{Wang2013}. Our study provides a new way to measure the spatial relative phase
between two light fields that have different frequencies. The
phase information is converted to intensity signals.
This can be used in phase-contrast microscopy.

We thank B. Gadway, J. Q. You and G. Juzeliunas for fruitful discussion. We acknowledge the support from the National Key Research and Development Program of China under Grant No.2018YFA0307200, the Joint Fund of National Natural Science Foundation of China 
(U1330203) and National Natural Science Foundation of China 
(No.91736209, 11574188).

 \bibliographystyle{nature}
\bibliography{sl1}

\begin{thebibliography}{10}

\bibitem{Klitzing1980}
Klitzing, K.~v., Dorda, G., and Pepper, M.
\newblock {\em Physical Review Letters}{ \bf 45}(6), 494--497 (1980).

\bibitem{Thouless1982}
Thouless, D.~J., Kohmoto, M., Nightingale, M.~P., and Dennijs, M.
\newblock {\em Physical Review Letters}{ \bf 49}(6), 405--408 (1982).

\bibitem{Kane2005}
Kane, C.~L. and Mele, E.~J.
\newblock {\em Physical Review Letters}{ \bf 95}(22), 226801 (2005).

\bibitem{Bernevig2006}
Bernevig, B.~A. and Zhang, S.-C.
\newblock {\em Physical Review Letters}{ \bf 96}(10), 106802 (2006).

\bibitem{Lohse2018}
Lohse, M., Schweizer, C., Price, H.~M., Zilberberg, O., and Bloch, I.
\newblock {\em Nature}{ \bf 553}, 55 (2018).

\bibitem{Zilberberg2018}
Zilberberg, O., Huang, S., Guglielmon, J., Wang, M., Chen, K.~P., Kraus, Y.~E.,
  and Rechtsman, M.~C.
\newblock {\em Nature}{ \bf 553}, 59 (2018).

\bibitem{HuegelParedes2014}
H{\"u}gel, D. and Paredes, B.
\newblock {\em Physical Review A}{ \bf 89}(2), 023619 (2014).

\bibitem{Mancini2015}
Mancini, M., Pagano, G., Cappellini, G., Livi, L., Rider, M., Catani, J., Sias,
  C., Zoller, P., Inguscio, M., Dalmonte, M., and Fallani, L.
\newblock {\em Science}{ \bf 349}(6255), 1510--1513 (2015).

\bibitem{Livi2016}
Livi, L.~F., Cappellini, G., Diem, M., Franchi, L., Clivati, C., Frittelli, M.,
  Levi, F., Calonico, D., Catani, J., Inguscio, M., and Fallani, L.
\newblock {\em Physical Review Letters}{ \bf 117}(22), 220401 (2016).

\bibitem{Atala2014}
Atala, M., Aidelsburger, M., Lohse, M., Barreiro, J.~T., Paredes, B., and
  Bloch, I.
\newblock {\em Nature Physics}{ \bf 10}(8), 588--593 (2014).

\bibitem{Stuhl2015}
Stuhl, B.~K., Lu, H.-I., Aycock, L.~M., Genkina, D., and Spielman, I.~B.
\newblock {\em Science}{ \bf 349}(6255), 1514--1518 (2015).

\bibitem{Wang2015a}
Wang, D.-W., Liu, R.-B., Zhu, S.-Y., and Scully, M.~O.
\newblock {\em Physical Review Letters}{ \bf 114}(4), 043602 (2015).

\bibitem{Wang2015b}
Wang, D.-W., Cai, H., Yuan, L., Zhu, S.-Y., and Liu, R.-B.
\newblock {\em Optica}{ \bf 2}(8), 712--715 (2015).

\bibitem{Chen2018}
Chen, L., Wang, P., Meng, Z., Huang, L., Cai, H., Wang, D.-W., Zhu, S.-Y., and
  Zhang, J.
\newblock {\em Physical Review Letters}{ \bf 120}, 193601 (2018).

\bibitem{Anisimovas2016}
Anisimovas, E., Ra{\v{c}}i{\=u}nas, M., Str{\"a}ter, C., Eckardt, A., Spielman,
  I., and Juzeli{\=u}nas, G.
\newblock {\em Physical Review A}{ \bf 94}(6), 063632 (2016).

\bibitem{Xu2018}
Xu, J., Gu, Q., and Mueller, E.~J.
\newblock {\em Physical Review Letters}{ \bf 120}(8), 085301 (2018).

\bibitem{An2018}
An, F.~A., Meier, E.~J., Ang’ong’a, J., and Gadway, B.
\newblock {\em Physical Review Letters}{ \bf 120}(4), 040407 (2018).

\bibitem{An2017}
An, F.~A., Meier, E.~J., and Gadway, B.
\newblock {\em Science Advances}{ \bf 3}(4) (2017).

\bibitem{Scully2006}
Scully, M.~O., Fry, E.~S., Ooi, C. H.~R., and Wódkiewicz, K.
\newblock {\em Physical Review Letters}{ \bf 96}(1), 010501 (2006).

\bibitem{Fleischhauer2000}
Fleischhauer, M. and Lukin, M.~D.
\newblock {\em Physical Review Letters}{ \bf 84}(22), 5094--5097 (2000).

\bibitem{Feldman1972}
Feldman, B.~J. and Feld, M.~S.
\newblock {\em Physical Review A}{ \bf 5}(2), 899.

\bibitem{Zhang2018}
{Zhang}, L., {Zhang}, L., {Niu}, S., and {Liu}, X.-J.
\newblock {\em ArXiv 1802.10061}{ \bf } (2018).

\bibitem{Rechtsman2013}
Rechtsman, M.~C., Zeuner, J.~M., Plotnik, Y., Lumer, Y., Podolsky, D., Dreisow,
  F., Nolte, S., Segev, M., and Szameit, A.
\newblock {\em Nature}{ \bf 496}(7444), 196--200 (2013).

\bibitem{Hafezi2013}
Hafezi, M., Mittal, S., Fan, J., Migdall, A., and Taylor, J.~M.
\newblock {\em Nature Photonics}{ \bf 7}, 1001 (2013).

\bibitem{Zhang2015}
Zhang, Y., Wu, Z., Belić, M.~R., Zheng, H., Wang, Z., Xiao, M., and Zhang, Y.
\newblock {\em Laser $\&$ Photonics Reviews}{ \bf 9}(3), 331--338 (2015).

\bibitem{Xue2017}
Xiao, L., Zhan, X., Bian, Z.~H., Wang, K.~K., Zhang, X., Wang, X.~P., Li, J.,
  Mochizuki, K., Kim, D., Kawakami, N., Yi, W., Obuse, H., Sanders, B.~C., and
  Xue, P.
\newblock {\em Nature Physics}{ \bf 13}(11), 1117--1123 (2017).

\bibitem{Bandres2018}
Bandres, M.~A., Wittek, S., Harari, G., Parto, M., Ren, J., Segev, M.,
  Christodoulides, D.~N., and Khajavikhan, M.
\newblock {\em Science}{ \bf }.
\newblock 10.1126/science.aar4005.

\bibitem{Haldane2008}
Haldane, F. D.~M. and Raghu, S.
\newblock {\em Physical Review Letters}{ \bf 100}(1), 013904 (2008).

\bibitem{Wu2016}
Wu, Z., Zhang, L., Sun, W., Xu, X.~T., Wang, B.~Z., Ji, S.~C., Deng, Y.~J.,
  Chen, S., Liu, X.~J., and Pan, J.~W.
\newblock {\em Science}{ \bf 354}(6308), 83--88 (2016).

\bibitem{Huang2016}
Huang, L., Meng, Z., Wang, P., Peng, P., Zhang, S.-L., Chen, L., Li, D., Zhou,
  Q., and Zhang, J.
\newblock {\em Nature Physics}{ \bf 12}, 540 (2016).

\bibitem{Haldane1988}
Haldane, F. D.~M.
\newblock {\em Physical Review Letters}{ \bf 61}(18), 2015--2018 (1988).

\bibitem{Jotzu2014}
Jotzu, G., Messer, M., Desbuquois, R., Lebrat, M., Uehlinger, T., Greif, D.,
  and Esslinger, T.
\newblock {\em Nature}{ \bf 515}(7526), 237--240 (2014).

\bibitem{Peyronel2012}
Peyronel, T., Firstenberg, O., Liang, Q.~Y., Hofferberth, S., Gorshkov, A.~V.,
  Pohl, T., Lukin, M.~D., and Vuletic, V.
\newblock {\em Nature}{ \bf 488}(7409), 57--60 (2012).

\bibitem{Ripka2016}
Ripka, F., Chen, Y.-H., Löw, R., and Pfau, T.
\newblock {\em Physical Review A}{ \bf 93}(5), 053429 (2016).

\bibitem{Greschner2017}
Greschner, S. and Vekua, T.
\newblock {\em Physical Review Letters}{ \bf 119}(7), 073401 (2017).

\bibitem{Lin2011b}
Lin, Z., Ramezani, H., Eichelkraut, T., Kottos, T., Cao, H., and
  Christodoulides, D.~N.
\newblock {\em Physical Review Letters}{ \bf 106}(21), 213901 (2011).

\bibitem{Regensburger2012}
Regensburger, A., Bersch, C., Miri, M.-A., Onishchukov, G., Christodoulides,
  D.~N., and Peschel, U.
\newblock {\em Nature}{ \bf 488}, 167 (2012).

\bibitem{Wu2014}
Wu, J.-H., Artoni, M., and La~Rocca, G.~C.
\newblock {\em Physical Review Letters}{ \bf 113}(12), 123004 (2014).

\bibitem{Huang2017}
Huang, Y., Shen, Y., Min, C., Fan, S., and Veronis, G.
\newblock {\em Nanophotonics}{ \bf 6}(5), 977 (2017).

\bibitem{Feng2017}
Feng, L., El-Ganainy, R., and Ge, L.
\newblock {\em Nature Photonics}{ \bf 11}(12), 752--762 (2017).

\bibitem{Wang2013}
Wang, D.-W., Zhou, H.-T., Guo, M.-J., Zhang, J.-X., Evers, J., and Zhu, S.-Y.
\newblock {\em Physical Review Letters}{ \bf 110}(9), 093901 (2013).

\end{thebibliography}

\section{methods}

\subsection{Experiment}

\begin{figure*}
\includegraphics[width=1.5\columnwidth]{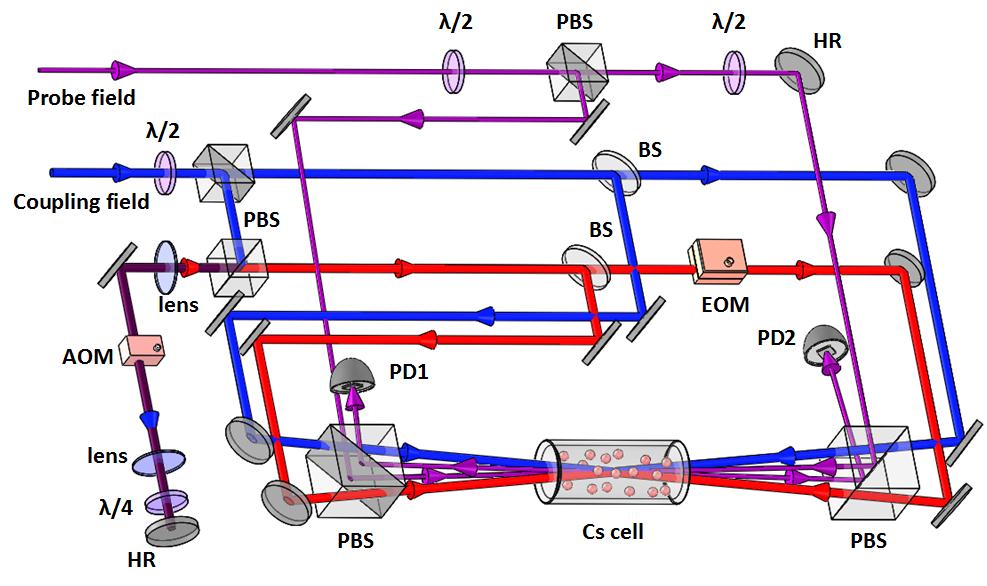}\caption{\textbf{The 
experimental set-up}. BS: beam splitter. PBS: polarization beam splitter. AOM: acousto-optical modulator.
EOM: elctro-optic modulator. HR: high reflection mirror. PD: photo detector. $\lambda/2$: half-wave plate. $\lambda/4$: quarter-wave plate.}
\label{method} 
\end{figure*}

The experimental set-up is schematically drawn in Fig. \ref{method}.
A Ti:Sapphire laser operating at $^{137}$Cs $D_{1}$ line (895 nm)
is split by a polarization beam splitter (PBS) into two beams for
the two coupling fields. The vertically polarized beam is detuned
by an acousto-optical modulator (AOM) in a double-pass configuration,
during which its polarization is rotated to the horizontal direction
by a quarter-wave plate. The two coupling fields are then split by
50/50 beam splitters (BS) to form two standing waves. The laser that
is detuned by the AOM serves as the near-resonant coupling field.
A plane wave component of the far-detuned standing wave goes through
an electro-optic modulator (EOM) to tune the spatial relative phase
$\phi/2$ between the two standing waves. Each plane wave component
of the near-resonant coupling field is aligned to overlap with the
counterpropagating component of the far-detuned coupling field. A
Toptica DL100 semiconductor laser with vertical polarization is split
by a 50/50 BS to generate two counter-propagating probe beams. The
probe fields and two coupling fields form a triangular configuration
along their propagation direction, as shown in Fig. \ref{fig1} (\textbf{B}).
The intersection angles between the probe field and the two coupling
fields are both 0.62 degree. The intersection angle between the two
coupling fields is 1.0 degree. These angles make sure the phase-matching
condition can be satisfied assisted by the dispersion of the atomic
gas. In addition, the non-degenerate wave-mixing signals are phase
matched in a different direction from the degenerate wave-mixing signals
that we detect. The $e^{-2}$ full width of the coupling field is
1.6 mm while the one for the probe field is 0.5 mm. The length of
the atomic vapour cell is 2 cm. The probe beam is totally covered
by the coupling beams in the atomic vapour. The cell length is small
enough that we consider the relative phase between the two standing
waves is a constant. The two reflected signals were simultaneously
recorded with photodiode detectors.

\subsection{Numerical simulation}

The reflectivities $R_{\pm}$ are obtained from the coupled-wave equations
\cite{Wang2015b}, 
\begin{equation}
\begin{aligned}\frac{\partial E_{p}}{\partial x} & =-\eta\text{Im}\beta_{0}E_{p}+i\eta\beta_{+1}e^{-i\Delta kx}E_{r},\\
\frac{\partial E_{r}}{\partial x} & =+\eta\text{Im}\beta_{0}E_{r}-i\eta\beta_{-1}e^{+i\Delta kx}E_{p},
\end{aligned}
\label{cwe}
\end{equation}
where $E_{p}$ and $E_{r}$ are the slowly-varying field amplitudes
of the probe field propagating in $+\hat{x}$ and the reflected field.
$\eta$ is a parameter relating the probability amplitudes to the
susceptibilities \cite{Wang2015b}. `$\text{Im}$' stands for the
imaginary part. $\Delta k=2({k}_{p}-{k}_{c})$ is the wavevector mismatch
in $\hat{x}$ direction. By setting the boundary condition $E_{p}(0)=E_{0}$,
$E_{r}(L)=0$ where $L$ is the length of the cell, we obtain $R_{+}=|E_{r}(0)/E_{0}|^{2}$
from the coupled wave equations. On the other hand, by setting $E_{p}(0)=0$,
$E_{r}(L)=E_{0}$, we obtain $R_{-}=|E_{r}(0)/E_{0}|^{2}$. In the
Supplementary Material, we show that the coupled-wave equation results
in the simple relation in Eq. (\ref{eq:ratio}).

In calculating $\beta_{j}$, we have used the master equation that
takes into account the decoherence and Doppler shifts of the atoms
(see Supplementary Material). We adopt a more rigorous approach based
on calculating the Fourier components of the atomic coherences in the
reference frame of moving atoms.

\end{document}